# Field-driven attosecond photoinjection dynamics in semiconductors

Giacomo Inzani[1], Lyudmyla Adamska[2], Amir Eskandari-asl[3], Nicola Di Palo[1], Gian Luca Dolso[1], Bruno Moio[1], Luciano Jacopo D'Onofrio[2], Alessio Lamperti[4], Alessandro Molle[4], Rocío Borrego-Varillas[5], Mauro Nisoli[1,5], Stefano Pittalis[2], Carlo Andrea Rozzi[2,‡], Adolfo Avella[3,6,7,†], Matteo Lucchini[1,5,*]

[1]Department of Physics, Politecnico di Milano, Piazza Leonardo da Vinci, 20133 Milano, Italy
[2]CNR - Istituto Nanoscienze, via Campi 213/A, I-41125 Modena, Italy.
[3]Dipartimento di Fisica "E. R. Caianiello", Università degli Studi di Salerno, I-84084 Fisciano, Italy.
[4]CNR-IMM, Unit of Agrate Brianza, 20864 Agrate Brianza, MB, Italy.
[5]Institute for Photonics and Nanotechnologies, IFN-CNR, 20133 Milano, Italy.
[6]CNR-SPIN, UoS di Salerno, I-84084 Fisciano (SA), Italy.
[7]Unità CNISM di Salerno, Università degli Studi di Salerno, I-84084 Fisciano (SA), Italy.

[‡]E-mail: carloandrea.rozzi@nano.cnr.it
[†]E-mail: a.avella@unisa.it
[*]E-mail: matteo.lucchini@polimi.it

**The route towards manipulation of the optoelectronic properties of matter beyond the current limits of electronics starts from a comprehensive study of the ultrafast dynamics triggered by interaction with light. Among them, a fundamental role is played by charge photoinjection, a complex process that stems from the interplay of many different physical phenomena, which cannot be easily disentangled. Single- and multi-photon absorption, diabatic tunnelling, intra-band motion, and field-driven band dressing, all concur in determining the overall excited electron population, dictating the electro-optical properties of a material. Here we investigate ultrafast photoinjection in a prototypical semiconductor (monocrystalline germanium) by using attosecond transient reflection spectroscopy. The precise pump-field characterization ensured by a simultaneous attosecond streaking experiment, in tandem with a comprehensive theoretical approach, allowed us to disentangle the different physical phenomena unfolding at different positions in the reciprocal space and at different timing within the envelope of the pump pulse. Moreover, we found that intra-band phenomena hinder charge injection, in contrast to what was previously observed for resonant, direct band-gap semiconductors.**

**Therefore, besides other known parameters as the central wavelength and peak intensity, our results indicate that the pulse temporal envelope and the local band structure probed by intra-band effects are of key importance to achieve an optimal control over the ultrafast carrier injection process and tailor the complex optical and electronic properties of a semiconductor on the few- to sub-femtosecond time scale.**

The possibility to control the physical properties of solids with light pulses is a fascinating goal at the core of many fields of technology and research. In this regard, the more recent developments of attosecond and few-femtosecond science have opened the possibility to study the very first instants of light-matter interaction and explore the ultimate speed limit at which charges can be modulated with light[1]. Pioneering experiments have shown that intense optical pulses with controlled waveform can be used to drive ultrafast charge dynamics and modulate the optical properties of the sample in the PHz range[2,3]. While this behaviour seems to be a rather ubiquitous phenomenon, the given physical explanation behind the ultrafast optical response varies with the optoelectrical properties of the material (e.g., band gap and electron effective mass), the pump laser intensity and central frequency. Light-field-induced electron tunnelling[4] and single- and multi-photon absorption[5,6], together with field-driven intra-band motion[7–9], all concur in defining the ultrafast optical properties. Under certain circumstances, one mechanism may dominate, allowing for the adoption of simplified approaches to obtain a physical interpretation of the experimental results[10]. Nevertheless, what happens in a generic material, whose global response cannot be reduced to a single physical phenomenon, or simply deduced by the behaviour of the high symmetry points, is not yet clear.

In this work, we employed an attosecond reflection spectroscopy scheme[11,12] to study the sub-cycle response of a prototypical semiconductor, germanium (Ge), under an intense ($I_0 \sim 8$ TW/cm²) infrared (IR) laser pulse. We observed ultrafast modulations of the sample reflectivity which oscillate at twice the pump frequency ($2\nu_{IR} \simeq 0.75$ PHz) and cannot be ascribed to a single physical mechanism, resorting to dimensionless parameters. Comparison with real-time time-dependent density functional theory (TDDFT) calculations and the results of a novel method, the dynamical projective operatorial approach (DPOA), allowed us to demonstrate that the observed features are strictly related to the carrier injection mechanism characterized by the complex interplay of different phenomena, whose relevance strongly depends on the timing within the pump pulse envelope and the exact region of the band structure involved. While a simplistic use of the Keldysh parameter[13] would suggest tunnelling at the high symmetry point of the band structure (Γ) to dominate, our results show that resonant multi-photon carrier injection away from Γ gives the largest contribution to the final excited electron population. Moreover, while single photon absorption dominates the first femtoseconds during light-matter interaction, two-photon absorption, tunnelling, intra-band motion, and band dressing at different points in the reciprocal space become relevant at later times, close to the maximum of the pulse envelope. The DPOA analysis, performed with a finer sampling of the reciprocal space, confirmed that while single-photon transitions characterize the optical response at low pump intensities, two-photon processes are dominant in the experimental conditions, and they are strongly influenced by the field-driven intra-band motion. In particular, we found intra-band motion to reduce the number of injected electrons, in clear contrast with what observed in semiconductors characterized by an almost resonant direct band gap[9].

By unveiling the ultrafast charge injection process in a realistic, complex, and multi-band scenario, our results widen the knowledge of strong-field phenomena in semiconductors and set the basis for their exploitation in many relevant research fields in optoelectronics and photonics, such as the realization of ultrafast optical switches[14,15], high-order harmonic generation in solids[16,17], and ultrafast modulators of extreme-ultraviolet (XUV) light for the next generation of UV-lithography[18,19], to name a few.

**Equilibrium and transient reflectivity of germanium**

Figure 1a shows a scheme of the pump-probe measurement. Further details on the experimental setup are given in the Methods, in Supplementary Section S1.1, and in Ref. [12]. An attosecond pulse

(~200 as) with spectrum centred around the Ge $M_{4,5}$ absorption edge ($3d_{5/2}$ ~29.2 eV, $3d_{3/2}$ ~29.8 eV), impinges onto the monocrystalline sample at the total external reflection angle[20] (66°). The unpumped reflectivity, $R_0(E)$, measured by collecting the spectral intensity of the attosecond radiation as a function of the photon energy $E$ (see Supplementary Section S1.5), is displayed in Fig. 1b, while the associated absorption coefficient $\alpha(E)$, obtained through a Kramers-Kronig analysis[21] (see Supplementary Section S1.6), is reported in Fig. 1c. Below the absorption edge, the reflectance of the sample is large: valence band (VB) states are filled, and XUV-induced transitions are forbidden. Above the absorption edge, a decrease in the reflectivity (increase in $\alpha(E)$) is associated with the onset of transitions from the $3d$ semi-core states to critical points of the conduction bands (CBs)[22]. A Fourier analysis of the post-absorption-edge region[23] and a multi-gaussian fit (see Supplementary Section S1.7), relate the details of the absorption coefficient to the unperturbed band structure (Fig. 1d), showing that, in the energy range of interest, the attosecond radiation is capable to probe mainly the three highest VBs and the four lowest CBs. The energy of the retrieved XUV-induced transitions is in good agreement with the published data[22,23].

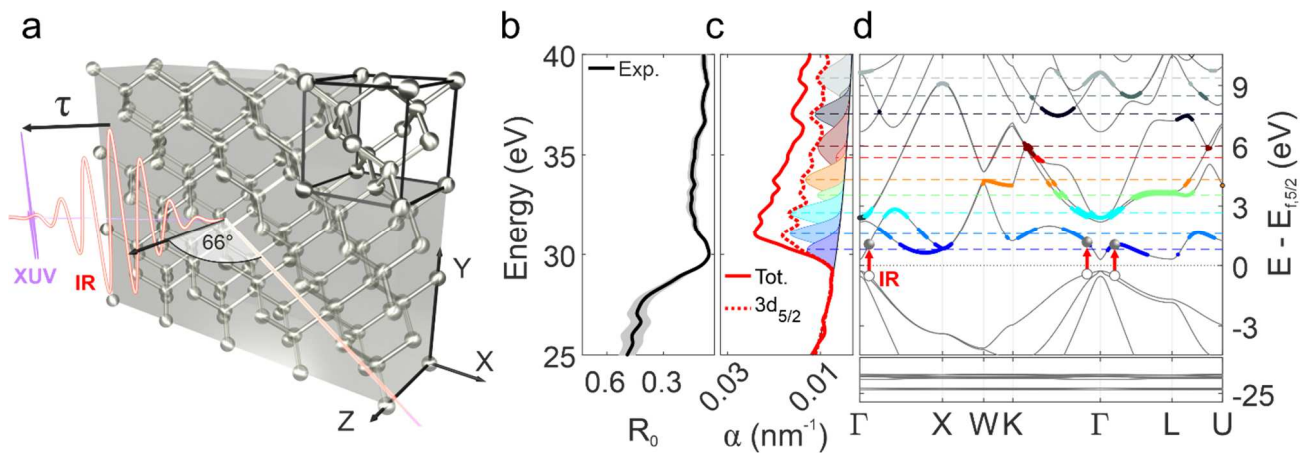

Fig. 1 | *Experimental scheme and static properties of germanium*. *a*, Schematic representation of the pump-probe measurement. *b*, Experimental Ge static reflectivity. The grey-shaded area covers twice the standard deviation of the average. *c*, Static absorption coefficient (red solid curve) and post-absorption-edge components associated with transitions from the $3d_{5/2}$ state to the CB (red dotted curve). A multi-gaussian fit (coloured Gaussian bells) allows for the extraction of the XUV-induced transitions. *d*, Equilibrium band structure of Ge. Coloured regions refer to transitions identified in panel (c). Red arrows identify the resonant IR single-photon transitions.

When an intense ($I_0 \sim 8$ TW/cm$^2$), 10-fs IR pulse with a central wavelength of about 800 nm, is used to pump the sample, it induces carrier injection from the VB to the CB and field effects which modify the sample reflectivity in the XUV range, $R_p$. Therefore, the ultrafast charge dynamics induced by the IR pump can be probed by the attosecond pulses by monitoring $R_p$ while changing the relative delay $\tau$ between the IR and XUV pulses. The experimental differential reflectivity, $\Delta R/R = [R_p(E,\tau) - R_0(E)]/R_0(E)$, measured for different values of the delay $\tau$ around the pump-probe temporal overlap, is represented in Fig. 2a together with the square of the pump vector potential $A_{IR}(\tau)$ (top panel, see Supplementary Section S1.3). The sample reflectivity is either increased (red regions) or decreased (blue regions), according to the XUV photon energy (Fig. 2b). Moreover, while few transient signals seem to follow the pump pulse envelope, vanishing when the interaction with the field is over (e.g., the lineout at 29.45 eV in Fig. 2b), most of the features survive for large positive delays, after the interaction with the pump pulse. Within the pump envelope, we observe additional fast oscillations which spread over the full energy range under consideration and that follow $A_{IR}^2(\tau)$.

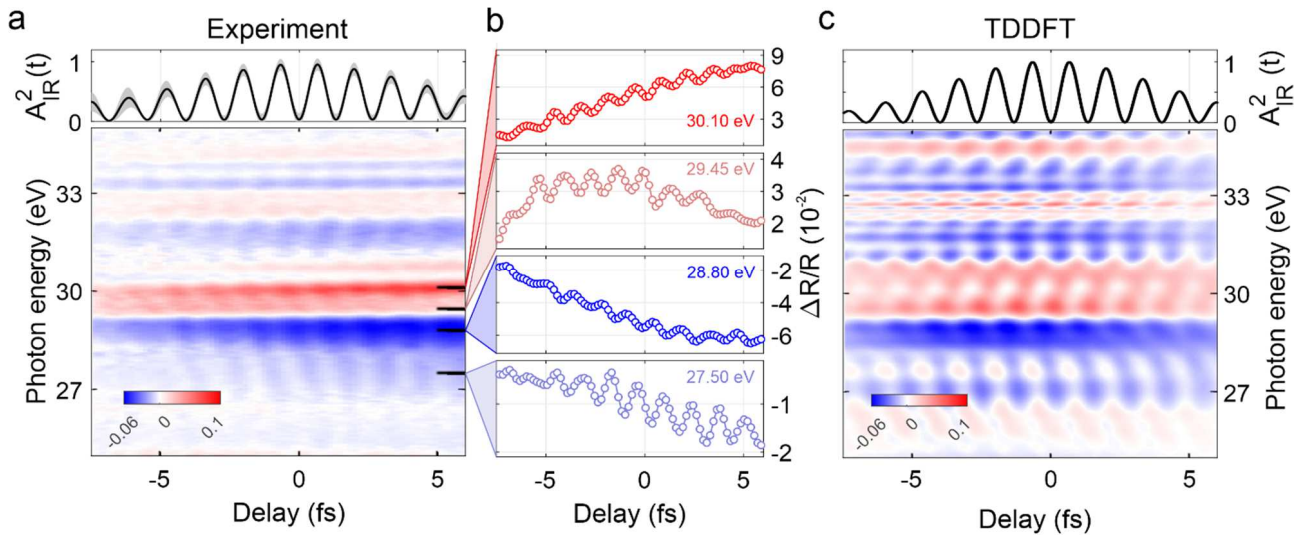

*Fig. 2* | **Attosecond transient reflectivity measurement in germanium.** *a, Experimental differential reflectivity trace, ΔR/R, for the Ge sample (main panel) and squared modulus of the simultaneously measured IR pump vector potential $A_{IR}^2(t)$ (top panel). **b**, Average ΔR/R calculated, over a 0.2-eV range, at the energy positions marked in a, around the material bandgap. **c**, ΔR/R calculated by TDDFT (main panel) and associated squared vector potential (top panel).*

This behaviour is similar to what previously observed in GaAs[9], suggesting that it may originate from a complex interplay between carrier injection from the VB to the CB[4,5], and IR-induced intra-band motion[8,11]. Despite using the same photon energy and comparable intensity, the considerably smaller energy gap at Γ in Ge (0.8 eV vs. 1.42 eV in GaAs) is expected to change the balance between resonant and non-resonant excitation, making a direct extension of the mechanisms found in GaAs not possible.

To gain further insight into the physical mechanism at play, we first performed real-time TDDFT calculations (see Methods and Supplementary Section S2.1), whose results are reported in Fig. 2c. The simulation results catch the qualitative behaviour of $\Delta R/R$, reproducing the regions of increased/decreased reflectivity and the $2\omega_{IR}$-oscillations. This allows us to study the phenomena ruling the photoinjection process during light-matter interaction. We will show that these phenomena can be disentangled by exploiting their different timing within the pump pulse envelope and the specific regions of the reciprocal space involved.

**Accessing light-matter interaction**

The Keldysh parameter[13] calculated for the direct energy gap at Γ ($E_g \sim 0.8$ eV)[24] gives a value of $\gamma \sim 0.37$, suggesting tunnel to play a dominant role[25]. Nevertheless, it is possible to identify specific regions of the reciprocal space in the proximity of Γ (red arrows in Fig. 1d), where the energy distance between VB and CB matches with the IR photon energy ($\hbar\omega_{IR} \sim 1.55$ eV). For such regions, we expect resonant one-IR-photon excitation to prevail[23]. In general, if the local energy gap is resonant with integer multiples of the IR photon energy and the corresponding matrix element is large enough, inter-band transitions associated with the absorption of multiple IR photons can occur. Besides direct carrier injection into the CB, the strong pumping field can have additional effects on the material. First, according to Bloch acceleration theorem, charges are accelerated in the reciprocal space along the polarization direction of the IR vector potential[10], a mechanism often referred to as intra-band motion[7]. Second, the pump field can dress the electronic bands, significantly changing their energies and thus affecting the related inter-band transitions. In this work, we will refer to these two effects together as intra-band contributions. These phenomena strongly depend on the local properties of the band structure, and all of them are expected to concur to the overall injection process. A more in-depth analysis of the excited electron population in the reciprocal space is thus needed to achieve further insights into the complex light-matter interaction under examination.

## Time-dependent k-resolved excited electron population

A way to disentangle the diverse physical mechanisms at play is to analyse how the population of excited electrons into the CBs evolves during the interaction with the IR pulse, at different points of the Brillouin zone. Figure 3a displays three families of k-points, grouped according to the ratio between the IR photon energy and the local energy gap, for which different mechanisms are expected to prevail: resonant single-photon absorption for those points where the energy gap $E_g$ is close to $\hbar\omega_{IR}$ (orange/violet dots), tunnel excitation at $\Gamma$ (blue dot), where $E_g < \hbar\omega_{IR}$, and two-photon absorption in those regions of the Brillouin zone where $E_g \sim 2\hbar\omega_{IR}$ (red/green dots).

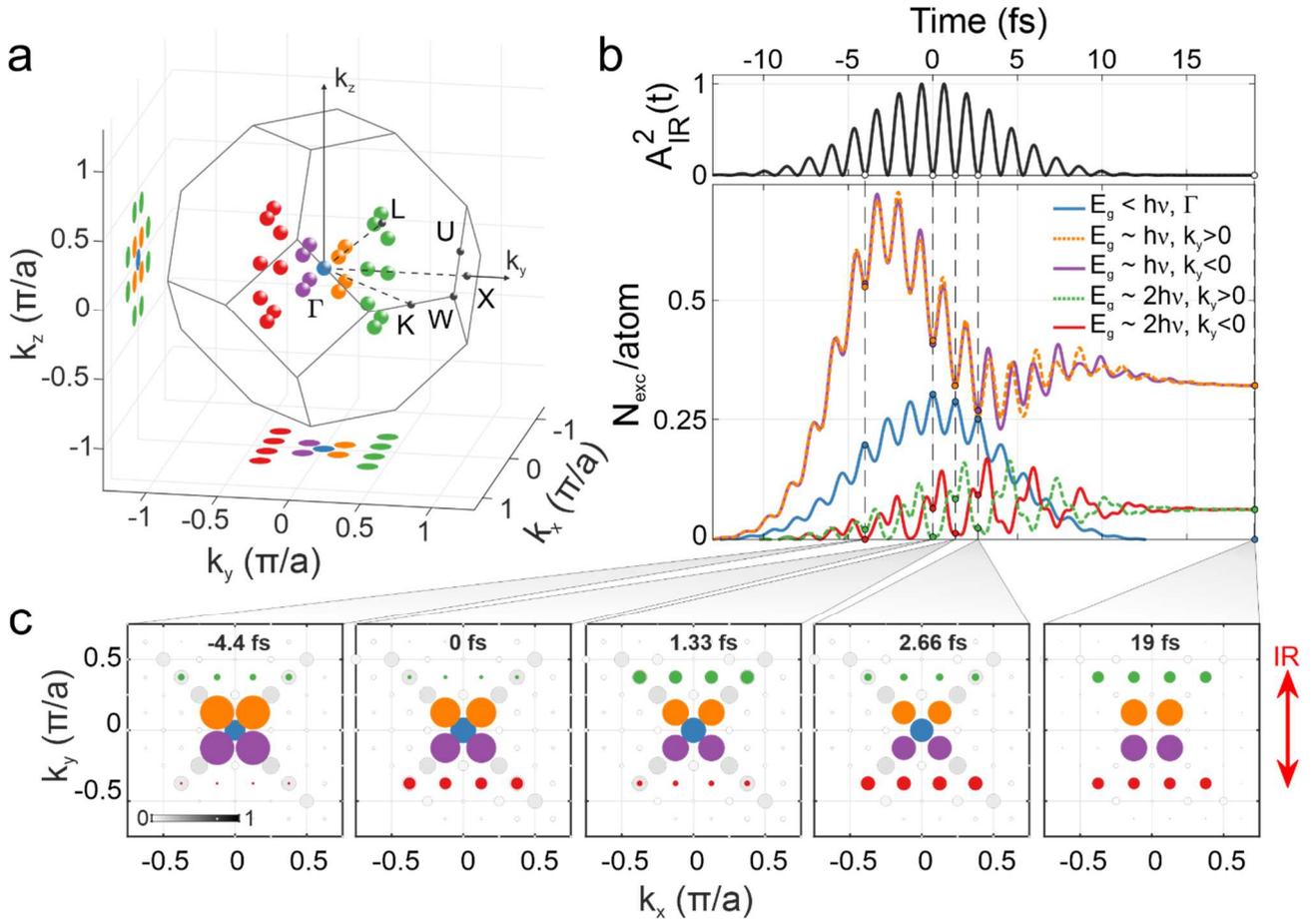

*Fig. 3 | **Time-dependent excited electron population in reciprocal space. a**, 3D location of the main k-points which participate in carrier injection. The blue dot is located at the $\Gamma$ point. Orange and violet dots represent those points where the energy gap between VB and CB is resonant with the IR photon energy. Red and green points are associated with a two-photon absorption mechanism. **b**, Time-dependent excited electron population computed by TDDFT at the selected k-points and normalised by the number of k points of the same family. Same colour coding as in Fig. 3a. **c**, Projection of the excited electron population along the $k_z$ direction at selected time instants marked by the dots in b. The size (and grey level) of each dot (same colour coding as in Fig. 3a) is proportional to the normalized excited electron population.*

The time evolution of the excited electron population calculated with TDDFT for a single k point of each of the above regions, is reported in Fig. 3b, while Fig. 3c shows its projection along the $k_z$ direction and for selected instants of time (see Supplementary Section S2.1.4 and S2.1.5 and Supplementary Animation SA1 for a full, time-dependent 3D representation in the reciprocal space). At negative times ($t = -4.00$ fs), the instantaneous IR intensity is low, and the dominating phenomenon is single-photon injection in the proximity of the Γ point (orange and violet dots/curves). Close to the peak of the pump pulse ($t = 0$ fs), where the IR intensity is maximum, single-photon injection is reduced in favour of diabatic tunnelling[10,25] across the direct bandgap at Γ (blue dot/curves, see Supplementary Section S2.2.3 and Supplementary Animation SA2). Two-photon absorption (red and green dots/curves) becomes relevant during the second half of the pump envelope (e.g., $t = 1.33$ and $2.66$ fs). In contrast with what is predicted by the Keldysh parameter, the residual electron population after the interaction with the pulse ($t = 19$ fs) is not given by tunnel ionization, but rather by single-photon and two-photon excitations. The population due to tunnelling at Γ instead follows the IR envelope, vanishing when the pulse is over.

As observed in the reflectivity data, all the populations are characterized by a fast oscillatory component that beats at twice the IR frequency, superimposed to a slower signal. After the peak of the pump pulse, the population of the 2-photon resonant points (green/red curves in Fig. 3b) oscillates out of phase, with a strong $\omega_{IR}$ component, partially visible also in the single-photon excitation (orange/violet curves in Fig. 3b). This effect can be explained by intra-band motion[7]. Since the pump polarization lies along the [010] crystallographic axis, electrons will be accelerated along the $k_y$ direction. Therefore, we expect intra-band motion to manifest as a mirror symmetry break along the $k_y$ axis. The population projections calculated with TDDFT and reported in Fig. 3c graphically display this phenomenon. At $t = 1.33$ fs, for instance, excited electrons significantly populate regions of the band structure associated with two-photon injection with $k_y > 0$ (green points), while their number in regions with the same bandgap but with $k_y < 0$ (red points) is negligible. After half a period of the pump field, at $t = 2.66$ fs, the situation is the opposite. It is important to notice that this population oscillation is not due to a direct electron transfer between $k$ points with positive and negative $k_y$ component. The maximum momentum change induced by the IR field ($eE_{IR}^{(max)}/\hbar\omega_{IR} \sim$ 0.8 nm$^{-1}$) is considerably smaller than the momentum separation between the red and green points in the reciprocal space (4.16 nm$^{-1}$). The origin of the $\omega_{IR}$-oscillations is thus different. As revealed through the DPOA analysis of the dynamical bands with and without intra-band contribution (see Supplementary Section S2.2.5), they come from the IR dressing the equilibrium bands and causing the local energy gaps to change in time. Since considering an opposite $k_y$ effectively corresponds to a change in the sign of the dressing field $A_{IR}(t)$, the $\omega_{IR}$ component of the red and green points oscillates out of phase.

**Dependence on the pump intensity**

Further insight can be obtained by studying the residual number of excited electrons left in the CB, $N_{res}$, as a function of the pump pulse peak intensity $I_{max}$ (see Supplementary Section S1.4). The $N_{res}$ calculated with TDDFT on an 8$^3$ grid in the $k$ space is reported in Fig. 4a by full-black circles. It scales as $\sim I_{max}^{1.443}$ (black dashed line) over a broad range between 0.5 and 20 TW/cm$^2$. Even though the Keldysh parameter $\gamma$ for the heavy-hole VB and first CB at Γ is smaller than 0.5 (see Supplementary Section S1.10), an $I_{max}^n$ power-law behaviour suggests that the overall charge injection mechanism is dominated by $n$-photon transitions[5,6] rather than diabatic tunnelling. This apparent contradiction and the fractional-exponent power law of $N_{exc}$, testify the inadequacy of a simplified two-band approach in describing real semiconductors[25]. In the case of Ge, an $1 \leq n \leq 2$ corresponds to multiphoton transitions dominating carrier injection, in agreement with our previous k-resolved analysis (Fig. 3b).

The complex physical scenario unravelled by our first analysis suggests that most of the carrier injection takes place away from the high symmetry points. To get a deeper insight and check the robustness of the proposed picture, we calculated $N_{exc}$ with DPOA (see Supplementary Section

S2.2.2). While the more common Houston state analysis fitted to first-principles calculations[7] has been used to decouple intra-band motion only for a limited number of bands[9,26,27], DPOA can distinguish between inter- and intra-band contributions for the whole band structure (see Supplementary Section S2.2.3). This provided us the opportunity to refine our identification of the relevant $k$ points, and better estimate the relative strength of the different physical mechanisms at play.

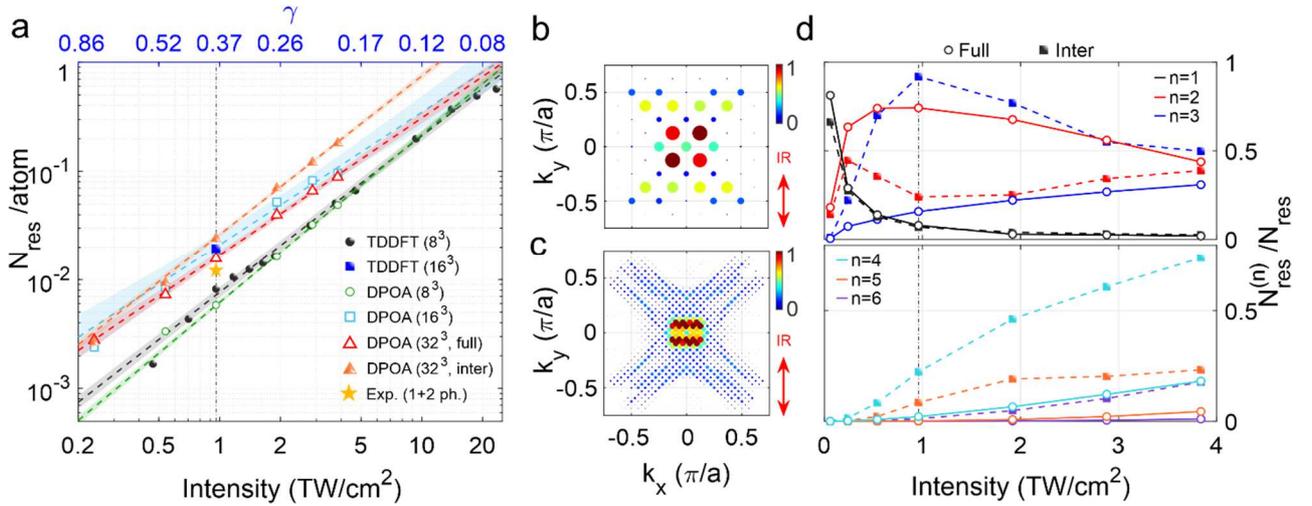

*Fig. 4| **Intensity dependence of the photoinjection process. a,** Total residual excited electron population, $N_{res}$, as a function of the pump peak intensity inside the sample computed by: TDDFT on a $8^3$ grid (full black circle), TDDFT on a $16^3$ grid (full blue square), DPOA on a $8^3$ grid (open green circles), DPOA on a $16^3$ grid (open light-blue squares), DPOA on a $32^3$ grid including (open red triangles) or excluding (full orange triangles) intra-band contributions. Values are compared with the expected residual population, computed considering only single- and two-photon processes (yellow star). Dashed lines represent linear fits in bi-logarithmic scales, while shaded regions extend over twice the fit confidence interval. The vertical black dash-dotted line marks the experimental condition. **b,c,** Projection of the excited electron population along the $k_z$ direction at $t = 19\,fs$ computed by DPOA on the $8^3$ and $32^3$ grid, respectively. The size and colour of each dot are proportional to the normalized excited electron population. **d,** Relative weight of n-photon transitions (different colours) including (open circles, solid curves) or excluding (full squares, dashed curves) intra-band contributions, as computed by DPOA on the $32^3$ grid. Each point is normalized to the associated $N_{res}$ computed by including intra-band contribu*tions.

The value of $N_{res}$ as a function of $I_{max}$, calculated with DPOA for a $k$ grid composed of $8^3$ points (same as TDDFT), is reported in Fig. 4a with open green circles. DPOA fairly reproduces the TDDFT results with $N_{res} \propto I_{max}^{1.548}$, and predicts residual population for the same relevant $k$ points (Fig. 4b, to be compared with the rightmost panel in Fig. 3c). Increasing the number of points to $16^3$, DPOA predicts a generally higher $N_{res}$, which scales as $I_{max}^{1.218}$ (Fig. 4a, light-blue open squares and dashed line). The higher value of $N_{res}$ obtained on a denser grid agrees with what calculated with TDDFT at the experimental intensity (Fig. 4a, full blue square), further validating the DPOA results. Further increasing the grid size to $32^3$, allows us to refine the scaling law for $N_{res}$ which is found to behave like $I_{max}^{1.260}$ on a $32^3$ grid (Fig. 4a, red open triangles and dashed line), proving that $1 \leq n \leq 2$ and the qualitative physical picture is unaffected by the grid size. Since the excited electron density is significantly larger than the Mott critical density[28] ($n_{Mott} \sim 4.68 \times 10^{14}$ cm$^{-3}$ vs. $6.91 \times 10^{20}$ cm$^{-3}$ predicted by DPOA on the $32^3$ grid, see Supplementary Sections S1.10), Coulomb correlations are strongly screened and the number of excited electrons can be evaluated by considering single- and two-photon absorption under the experimental conditions (see Methods and Supplementary Section S1.9). The result is marked with a yellow star in Fig. 4a and fairly compares with the calculated $N_{res}$. The finer $k$ grid can also be used to refine the identification of the points characterized by a residual electron population after interaction with the pump pulse, which appear to be denser around Γ (Fig. 4c) when compared with the $8^3$ results (Fig. 4b). To better investigate the relative importance of the different excitation mechanisms at play, using DPOA, we calculated the residual excited electron population associated to each $n$-photon transition, $N_{res}^{(n)}$, in a narrower region around the experimental $I_{max}$ (see Supplementary Section S2.2.4). Figure 4d presents the results including (open markers) and excluding (full markers) intra-band contributions, normalized by the associated $N_{res}$ where all contributions are included. While in the low-intensity limit single-photon transitions (black open circles) prevail, at the experimental conditions the dominant mechanism is represented by two-photon processes (red open squares). Further increasing the pump intensity, three- and four-photon contributions become relevant.

The results of Fig. 4d clearly show that intra-band phenomena strongly reshape the relative weight of the different $n$-photon processes. At the experimental pump intensity, multiphoton transitions involving more than three photons are less relevant than two-photon processes (compare full and open markers in Fig. 4d). As a result, $N_{res}$ grows quicker with the intensity if only inter-band transitions are considered ($\sim I_{max}^{1.454}$, orange full triangles and dashed line in Fig. 4a). Moreover, intra-band effects are found to generally reduce the number of excited electrons by the pump pulse, in clear contrast to what previously observed in GaAs[7,9]. Due to the wider and resonant energy gap (1.42 eV vs. 0.8 eV), in the case of GaAs inter-band transitions are mostly mediated by single-photon absorption at Γ where the energy bands have a local maximum/minimum. In Ge, Fig. 4c shows that charge injection happens around Γ in a region where CB and VB are significantly dispersive (red arrows in Fig. 1d). For this reason, their energy spacing is more affected by the IR field dressing, which can easily move relevant regions of the band structure in/out of resonance, changing the relative weight of different multiphoton processes, with a net effect of decreasing the total residual charge carrier density.

The injection of carriers by ultrashort and intense laser pulses from the VB to the CB of a semiconductor is therefore a complex process, that cannot be explained with a simplified, intuitive description. Indeed, even if a simplistic use of the Keldysh parameter would suggest diabatic tunnelling to dominate carrier injection, we found that multi-photon transitions, intra-band motion and band dressing strongly concur to the overall electronic excitation in a real, multi-band semiconductor like Ge. By combining the sub-cycle time resolution of the experimental results with advanced theoretical calculations, we found that the different mechanisms exhibit a different timing within the pump pulse envelope and thus can be partially disentangled. During the first few femtoseconds under the pulse leading edge, when the instantaneous intensity is relatively weak, single-photon excitation dominates. Closer to the pulse peak, diabatic tunnelling and, a few-fs later, multi-photon absorption become relevant. While the former does not contribute to the total residual electron population, the latter is responsible for most of it. Further theoretical analysis revealed a non-trivial dependence on the pulse intensity, deeply affected by field-dressing effects. In contrast with what previously reported, intra-band contributions reduce the injected charge carrier density by decreasing the probability of simultaneous absorption of more than three pump photons. Our study thus suggests that the actual role of intra-band phenomena strongly depends on the ratio between the pump photon energy and the local energy gaps of the material, which determine the exact location of the multi-photon resonant points where charge injection occurs. As intra-band motion is also influenced by the actual band structure in the proximity of the injection points, the net optical power needed to reach the desired carrier injection is expected to vary with the relative orientation between the pump field and the crystal lattice. The findings reported in this work hence suggest that the optimal control of charges with light in a semiconductor can be achieved not only by selecting the correct pulse wavelength and intensity, but also by tailoring its temporal properties to favour/inhibit a certain injection mechanism, and by selecting the crystal/polarization orientation that enables exploitation of intra-band effects.

By shedding new light onto such a fundamental and yet complex physical phenomenon, our work moves another important step towards achieving full optical control over electron excitation in semiconductors, a needful prerequisite to enable the manipulation of the electronic and optical properties of matter with light on the few- to sub-fs time scale[29–31], well before any possible intra- or inter-valley relaxation process has occurred[20,32]. Future experiments conducted on compounds with controlled energy gap or with tunable pump pulse wavelength will determine whether there is an intrinsic limit to the net amount of information per exchanged power that can be written in matter with light.

## Methods

### Sample.

The sample is a commercial 2-mm thick intrinsic (undoped) portion of a germanium wafer by Active Business Company GmbH[33]. The resistivity of the sample is 40 Ω·cm from the manufacturer specifics. The surface of the sample is (001) and it has optical quality. The electric field of pump and probe pulses is along the [010] crystallographic direction. To remove the native oxide layer, the sample was cleaned in an ultrasonic bath with deionized water and then dipped in a 10% hydrofluoric acid (HF) solution[34]. After cleaning, the sample was stored and measured in high vacuum conditions ($p < 1 \times 10^{-6}$ mbar) to avoid oxidation and contamination. Further details are presented in Supplementary Section S1.2.

**Experimental setup.**

The experimental setup was extensively described elsewhere[12] (see Supplementary Section S1.1). It consists of a commercial Ti:sapphire laser system followed by a hollow core fiber compressor[35], that delivers 10-fs, 800-nm 1-mJ light pulses at a 10-kHz repetition rate. 70% of the input pulse energy is focused on an argon gas target to generate isolated XUV attosecond pulses (central energy ~30 eV) in the high-order harmonic generation process with the double-optical gating technique[36]. The remaining fraction of the input energy is further reduced to 10 µJ with a motorized iris and suitably delayed, forming pump pulses. A shutter (50% duty cycle at 1 Hz) blocks the pump beam, allowing to acquire probe-only measurements. Pump and probe pulses are collinearly recombined via a drilled mirror and the relative pump-probe delay is actively stabilized. A first toroidal mirror focuses both pulses onto an argon gas target, where an attosecond streaking experiment[37] is performed. A nonlinear fit allows for the precise calibration of the pump-probe delay and characterization of the pump and probe pulses (see Supplementary Section S1.3). The diverging radiation is refocused by a second toroidal mirror onto a germanium sample at 66° of incidence[20]. Both pump and probe pulses are linearly and s-polarized with respect to the sample surface. The XUV radiation reflected by the sample is steered by a gold plane mirror into an XUV spectrometer. The XUV spectra are collected by a micro-channel plate, a phosphorous screen, and a camera.

**Equilibrium reflectance measurement.**

To measure the equilibrium reflectance, the pump beam was blocked. We measured the XUV photon spectrum as it is reflected by the germanium sample, $I_{Ge}(E)$, and by a reference gold mirror, $I_{Au}(E)$, in the same conditions as a function of the XUV photon energy $E$. Computing the reflectance of gold, $R_{Au}(E)$, at the chosen incidence angle from published data[38], we can obtain the reflectance of the germanium sample as $R_{Ge}(E) = I_{Ge}(E) \cdot \frac{R_{Au}(E)}{I_{Au}(E)}$. The results in Fig. 1b are the average of 30 independent measurements. For further details on the analysis, see Supplementary Section S1.5.

**Attosecond transient reflectance measurement.**

In an attosecond transient reflectance measurement, we acquire XUV photon spectra reflected by the germanium sample when the IR pump is present, $I_p(E,\tau)$, or blocked, $I_0(E)$. From these quantities, it is possible to compute the differential reflectivity of the sample as:

$$\Delta R/R\,(E,\tau) = [R_p(E,\tau) - R_0(E)]/R_0(E) = [I_p(E,\tau) - I_0(E)]/I_0(E) \qquad (1)$$

100 individual XUV spectra were acquired in each shutter condition for each delay value. For each measurement, the voltage of the microchannel plate and phosphorous screen and the integration time of the camera was set to fully exploit the dynamic range of the camera. The edge-pixel referencing method[39] was applied to each measurement to remove correlated noise coming from the XUV spectrum fluctuations (see Supplementary Section S1.8). The delay axis of each measurement was independently calibrated using the simultaneous attosecond streaking measurement. The results in Figs. 2a, b are the average of 5 independent measurements (see Supplementary Section S1.8).

**Experimental number of excited electrons.**

The number of excited electrons per Ge atom in experimental conditions is estimated starting from the measured spectral power and pulse energy, accounting for transmission at the surface. During propagation through the sample, the intensity of the IR radiation, $I_{IR}$, changes due to single- and two-photon absorption[40] as $\frac{\partial I_{IR}}{\partial z} = -\alpha I_{IR} - \beta I_{IR}^2$, where $\alpha$ and $\beta$ are, respectively, the single- and two-photon absorption coefficients. The energy-dependent linear absorption coefficient $\alpha$ is reported in literature[41], while a constant two-photon absorption coefficient $\beta = 1500$ cm/GW was estimated[42]. The general solution to this equation is

$$I_{IR}(z) = -\frac{\alpha I_{IR}^{(0)}}{\beta I_{IR}^{(0)} - \left(\alpha + \beta I_{IR}^{(0)}\right)e^{\alpha z}}, \qquad (2)$$

where $I_{IR}^{(0)}$ is the IR spectral intensity immediately after the sample surface. For $\beta = 0$, this equation gives the usual expression $e^{-\alpha z}$ for single-photon absorption. The effective length probed by the XUV radiation is assumed to be $5d_x$, where $d_x \sim 50$ nm is the estimated penetration length of the XUV radiation inside the sample. The fraction of absorbed photons can be written as:

$$\eta_{ph}(z) = \frac{I_{IR}^{(0)} - I_{IR}(z)}{I_{IR}^{(0)}} = \frac{\left(\alpha + \beta I_{IR}^{(0)}\right) \cdot (1 - e^{\alpha z})}{\beta I_{IR}^{(0)}(1 - e^{\alpha z}) - \alpha e^{\alpha z}}. \tag{3}$$

The absorbed number of photons is finally given by multiplying $\eta_{ph}$ by the incident number of photons. For $\beta = 0$ (single-photon absorption), the number of excited electrons $n_{el}^{(1)}$ is equal to the number of absorbed photons $n_{ph}^{(1)}$. For $\beta \neq 0$ (single- and two-photon absorption), the increment in the number of absorbed photons, $n_{ph}^{(1+2)} - n_{ph}^{(1)}$, that is only related to two-photon absorption, must be divided by 2 to obtain the additional number of excited electrons, $n_{el}^{(2)} = \left(n_{ph}^{(1+2)} - n_{ph}^{(1)}\right)/2$. The total number of excited electrons in the combined single- and two-photon absorption case is thus $n_{el}^{(1+2)} = n_{el}^{(1)} + n_{el}^{(2)}$. Dividing these quantities by the volume pumped by the IR pulse and probed by the XUV radiation, we estimate of the excited electron density in the sample. For further details, refer to Supplementary Section S1.9.

**Time-dependent density functional theory.**

TDDFT calculations were performed using the Elk Software Suite[43,44], which uses an all-electron full-potential linearized augmented-plane wave numerical representation of Bloch orbitals in extended periodic systems, avoiding the pseudopotential approximation. To open the gap in the band structure of germanium, we used the PBE+U+J functional[45,46]. The full Brillouin zone was sampled on a $8 \times 8 \times 8$ grid. Real-time evolution was simulated with 1.2-as time step for almost 40 fs. The shape and intensity of pump and probe pulses were fitted to the experimental ones. Real-time evolution of the system subject to pump and probe pulses yields the time-dependent current, which is Fourier transformed to obtain the complex dielectric function. The latter is used to calculate reflectance for each relative delay of pump and probe pulses. This data is assembled as a heatmap to generate the simulated transient differential reflectance image. Lastly, we analyse the time-dependent projected occupations from the pump-only time-evolution. The detailed description can be found in the Supplementary Section S2.1.

**Dynamical projective operatorial approach.**

The Dynamical Projective Operatorial Approach (DPOA) exploits (i) the possibility of using sets of composite operators[47–51] to close the hierarchy of the equations of motion (EOM) of the operators relevant to the description of the system under analysis and (ii) that the linearity of the EOM achieved through this procedure is preserved once one describes the application of an electromagnetic pulse to the system within the dipole gauge of the minimal coupling approach[52]. The overall linearity permits to project exactly the time-dependent operators onto their equilibrium counterparts and to devise very effective approximations truncating the hierarchy of EOM or selecting the operators in the operatorial basis set according to the wanted properties and/or responses of the system under analysis. To access larger and larger grids in momentum, in the present work we neglected the dynamical contribution of the Coulomb interaction (i.e., we kept only the single-particle term of the Hamiltonian) and used values for the hopping and dipole matrices coming from a Wannierization of the DFT results (including the static contribution of the Coulomb interaction). Once the projection matrices are computed within DPOA (see Supplementary Section S2.2 for more details), one can compute the band populations and, within the linear response theory, the optical properties of the sample.

**Acknowledgments**

This project has received funding from the European Research Council (ERC) under the European Union's Horizon 2020 research and innovation programme (grant agreement No. 848411 title AuDACE), from MIUR PRIN aSTAR, Grant No. 2017RKWTMY and Laserlab-Europe EU-H2020 GA no. 871124.

**Author Contributions**

G.I, L.A. and A.E. contributed equally to this work. M.L., M.N. and R.B.V. conceived the experiment. G.I, N.D.P, G.L.D and B.M. performed the measurements and contributed to the definition of the experimental procedures. G.I. evaluated and analysed the results, calculated the sample absorption, and estimated the experimental excited electron density. A.L. and A.M. provided the samples. L.A., S.P. and C.A.R. designed, performed, and analysed DFT and TDDFT simulations. L.D. and A.A. performed initial DFT calculations. A.E. and A.A. developed DPOA, computed the dynamical bands and evaluated the inter-/intra-band contributions. L.D. and A.E. computed electron populations in DPOA. All authors participated in the scientific discussion. G.I. and M.L. wrote the first version of the manuscript to which all authors contributed.

**Competing interests**

The authors declare no competing interests

**Data availability statement**

All data generated and analysed during this study are available from the corresponding author upon reasonable request.

**Code availability statement**

All the custom codes used in this study are available from the corresponding author upon reasonable request.